\documentclass[manuscript]{acmart}
\usepackage{booktabs}
\usepackage{longtable}
\usepackage{csquotes}
\usepackage{comment}
\usepackage{tabularx}
\usepackage{fontawesome}
\usepackage{marvosym}
\usepackage[utf8]{inputenc}
\usepackage{color,soul}
\usepackage{xspace}
\usepackage{hyperref}
\usepackage{environ}
\usepackage{ifthen}
\usepackage{xcolor}

\definecolor{lightpurple}{HTML}{DCD6F7}
\definecolor{midpurple}{HTML}{A6B1E1}
\definecolor{darkpurple}{HTML}{3D3B8E}
\setulcolor{midpurple}

\newboolean{colortoggle}
\setboolean{colortoggle}{true} 

\newcommand{\ud}[1]{%
    \ifthenelse{\boolean{colortoggle}}{\textcolor{black}{#1}}{#1}%
}

\AtBeginDocument{%
  \providecommand\BibTeX{{%
    \normalfont B\kern-0.5em{\scshape i\kern-0.25em b}\kern-0.8em\TeX}}}

\setcopyright{acmcopyright}
\copyrightyear{2024}
\acmYear{2024}
\acmDOI{XXXXXXX}

\acmConference[CSCW '24]{Make sure to enter the correct
  conference title from your rights confirmation email}{November 09--13,
  2024}{San José, Costa Rica}
\acmPrice{15.00}
\acmISBN{978-1-4503-XXXX-X/18/06}

\author{Li Qiwei}
\affiliation{
  \institution{University of Michigan}
  \country{USA}
}
\email{author@example.com}

\author{Allison McDonald}
\affiliation{
  \institution{Boston University}
  \country{USA}
}
\email{author@example.com}

\author{Oliver L. Haimson}
\affiliation{
  \institution{University of Michigan}
  \country{USA}
}
\email{author@example.com}

\author{Sarita Schoenebeck}
\affiliation{
  \institution{University of Michigan}
  \country{USA}
}
\email{author@example.com}

\author{Eric Gilbert}
\affiliation{
  \institution{University of Michigan}
  \country{USA}
}
\email{author@example.com}

\begin{document}

\color{black}
\title[Sociotechnical Stack: Opportunities for NCIM Research]{\color{black}The Sociotechnical Stack: Opportunities for Social Computing Research in Non-consensual Intimate Media}

\newcommand{\sts}{\emph{sociotechnical stack}}

\renewcommand{\shortauthors}{Qiwei, et al.}

\begin{abstract}

\vspace{-0.7cm}

\begin{figure}[ht]
    \centering
    \includegraphics[width=\linewidth, trim={50 200 800 -80}, clip]{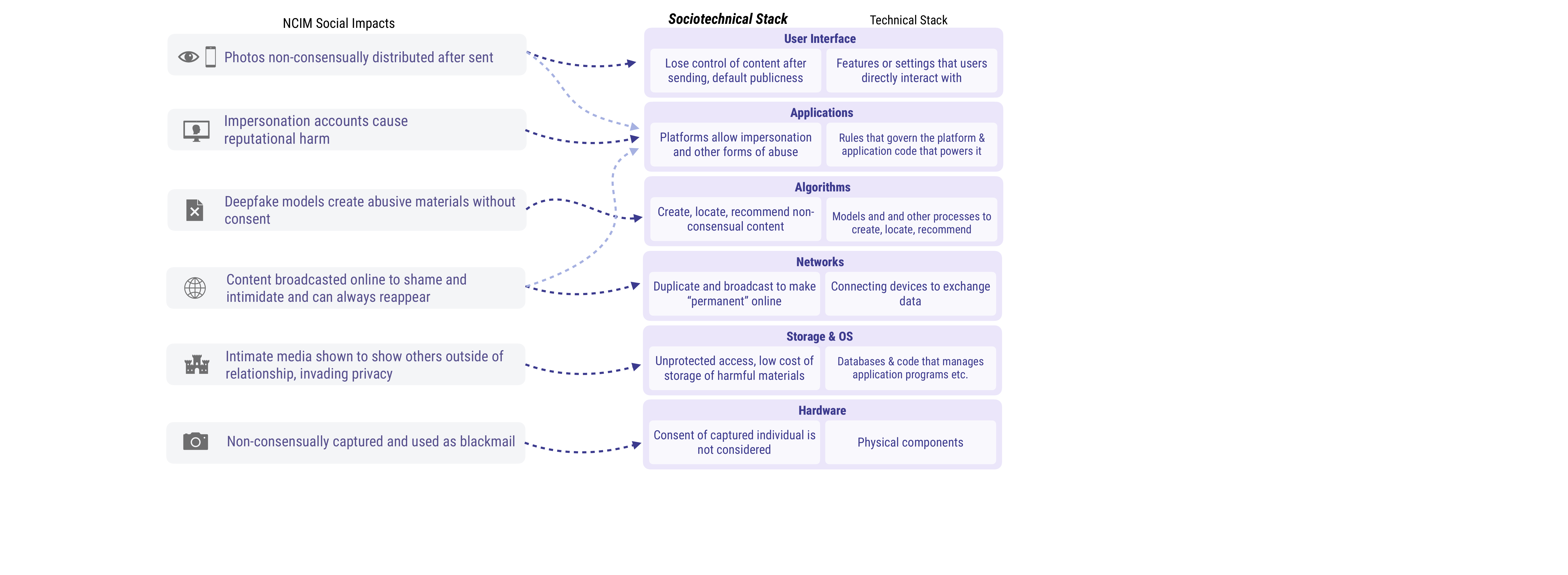}
    \caption{The \emph{sociotechnical stack} is a conceptual framework for analyzing sociotechnical problems. It maps social impacts (left) to layers of a traditional technical stack (right). Here, non-consensual intimate media (NCIM) social impacts are analyzed and mapped to corresponding layers in the technical stack. Darker arrows indicate a more direct connection from social impact to stack layer.}
    \label{fig:sstfig}
\end{figure}

\vspace{-0.3cm}

Non-consensual intimate media (NCIM) involves sharing intimate content without the depicted person's consent, including ``revenge porn'' and sexually explicit deepfakes. While NCIM has received attention in legal, psychological, and communication fields over the past decade, it is not sufficiently addressed in computing scholarship. This paper addresses this gap by linking NCIM harms to the specific technological components that facilitate them. We introduce the \sts{}, a conceptual framework designed to map the technical stack to its corresponding social impacts. The sociotechnical stack allows us to analyze sociotechnical problems like NCIM, and points toward opportunities for computing research. We propose a research roadmap for computing and social computing communities to deter NCIM perpetration and support victim-survivors\footnote{Our choice of the term \textit{victim-survivor} reflects that someone has been violated (victim), but is recovering or has recovered (survivor). We adopt these terms and definitions from sexual-abuse activism communities.} through building and rebuilding technologies.

\end{abstract}



\received{July 2023}
\received[revised]{April 2024}
\received[accepted]{July 2024}

\maketitle

\vspace{-0.5em}

\section{Introduction} \label{motivation}

\textit{\textcolor{purple}{Content warning: This paper talks about online sexual abuse.}}

\begin{figure}[h]
    \centering
    \includegraphics[width=\linewidth, trim={600 290 1050 200}, clip]{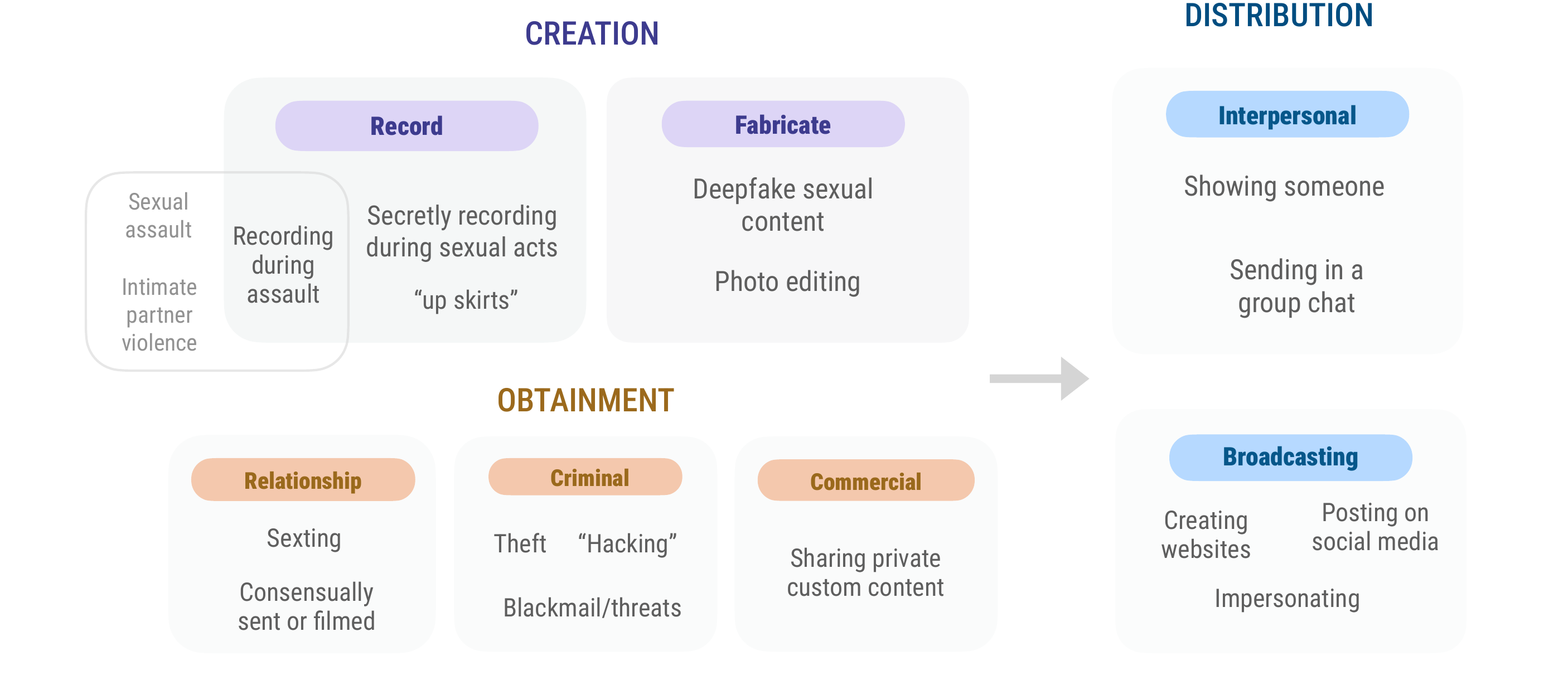}
    \caption{Non-consensual intimate media (NCIM) creation, obtainment, and distribution.}
    \label{fig:hero}
\end{figure}

\noindent
Non-consensual intimate media (NCIM) is the unauthorized creation, sharing, or distribution of sexual content containing someone's body or likeness. This form of online abuse usually targets women, who constitute about 90\% of all cases~\cite{eaton20172017,ccri2014revenge,safedigitalintimacy}. NCIM cases can be varied, ranging from sexually explicit deepfakes, to covert recordings, to sextortion (See Fig. \ref{fig:hero}). NCIM is common\textemdash studies suggest 1 in 6 adults have had their intimate content shared without consent~\cite{powell2024multi, ruvalcaba2020nonconsensual}. NCIM is also deeply traumatic---almost all victim-survivors experience severe anxiety and depression. About half seriously consider ending their own lives, and in some cases, women and girls commit suicide to escape emotional pain and social shame.~\cite{compton2024deepfake,ccri2014revenge,eaton20172017}. While there is widespread agreement that NCIM is harmful and some agreement that it should be regulated, there is less consensus on \textit{how} to do so~\cite{citron2014criminalizing,de_angeli_reporting_2023,citron2020internet,franks_drafting_2014}. The law is evolving to recognize NCIM, but access to legal recourse is limited and cannot prevent continued re-distribution. Victim-survivors may request content removal directly from platforms, but these efforts are typically laborious, emotionally taxing, and frequently futile due to the diverse jurisdictions governing online platforms~\cite{mcglynn_its_2021,de_angeli_reporting_2023}.

Effectively addressing NCIM requires both sociotechnical and computing research. Sexual abuse and patriarchal socio-historical conditions have enabled power and domination of others, typically women, via sex and intimacy. However, new technology has weaponized this at an unprecedented scale.

This paper highlights the essential links between NCIM and its enabling technical components. Drawing on existing surveys and interviews with more than 400 victim-survivors, we identify a critical gap: technologies, including social media platforms, file formats, and content recommendation algorithms, do not merely enable but actively \textit{create} NCIM at scale. To systemically analyze this, we introduce a conceptual framework called the \sts{}. In software engineering and computer science, a ``tech stack'' refers to a layered structure of technologies used to build systems, like the TCP/IP protocols in computing networking. We extend this concept to include \textit{social impacts}, integrating the technical and social dimensions of systems. Our analysis shows how certain layers, such as a content recommendation algorithm, can unintentionally promote NCIM. The sociotechnical stack illustrates the ways our digital environment facilitates risks and harms for NCIM victim-survivors, including the non-consensual creation of sexual content via generative AI and unauthorized redistribution of privately shared intimate media. We propose a roadmap for sociotechnical research to address NCIM and alleviate risks for victim-survivors.

\noindent
This paper's main contributions are:

\begin{itemize}
    \item \textbf{\textit{The sociotechnical stack conceptual framework:}} Intended for analyzing social computing issues, the sociotechnical stack framework breaks down complex sociotechnical problems into specific technical components that contribute to social impacts and harms.
    \item \textbf{\textit{A sociotechnical analysis of NCIM:}} This analysis sheds light on how various elements of the technical stack facilitate NCIM. We identify gaps in the current research on NCIM and demonstrate how adopting a sociotechnical perspective can help address these issues. 
    \item \textbf{\textit{A research agenda for NCIM:}} Finally, we propose important areas for further computing research on NCIM, at different layers of the sociotechnical stack.
\end{itemize}

\section{Overview of NCIM} \label{problemspace}

The essence of NCIM revolves around the concept of consent---or its absence. Defined as permission to engage in specific activities, consent in sexual contexts means agreeing to engage in particular sexual acts. However, consent transcends mere agreement; it is a dynamic practice influenced by societal norms and power dynamics ~\cite{srinivasan2021right}. The concept of consent, both within and beyond sociotechnical studies, is complex, nuanced, and frequently debated. The debate extends into how consent is obtained, maintained, and revoked; for example, affirmative consent requires ongoing, explicit agreement from all parties, emphasizing a ``yes means yes'' standard over the traditional ``no means no''~\cite{tuerkheimer2015affirmative,halley_move_2016}. However, this approach has been criticized for not capturing non-verbal cues and ignoring the complexities of sexual power dynamics ~\cite{jozkowski2015barriers,torenz2021politics}. 

Violations of consent can happen at many points in the lifecycle of NCIM. Critical moments in NCIM include content \textit{creation} (e.g., filming through covert cameras or using generative AI models to ``nudify'' someone), \textit{obtainment} (e.g., with media initially sent during intimate messaging), and \textit{distribution} (e.g., unauthorized sharing via direct messages or public online platforms) (see Fig. \ref{fig:hero}). Even content initially shared consensually within private contexts can lead to NCIM if shared beyond intended bounds. The impact ranges from isolated incidents to widespread distribution, affecting victim-survivors differently. For instance, sharing a photo with a friend may breach privacy, though without the intent to harm the subject, while uploading to a ``revenge porn'' site is a deliberate act of injury.

NCIM's prevalence is extensive and likely underreported, compounded by technical challenges in detection and social stigma that deter reporting. One 2017 survey found that 5\% of respondents admitted to non-consensually sharing intimate photos~\cite{eaton20172017}. \ud{This almost certainly undercounts people who consume NCIM online. A }2014 study estimated around 11,000 websites hosted non-consensual content---a figure that has likely grown~\cite{citron2014hate,ccri2014revenge,eaton20172017}. \ud{These sites range in character and include those dedicated to photos of ``up skirts'' (photos taken up a person's legs in public); contests (e.g., which user-uploaded photos get the most likes); trades (treating victim-survivors like baseball cards); or solicitations (in which users ask for photos of someone); and many others~\cite{citron2022fight,mcglynn_its_2021,eaton20172017}. }

The consequences of NCIM are severe, and a significant majority of victims experience emotional and social repercussions: 93\% of NCIM victims experience considerable emotional distress, 82\% encounter disruptions in their social and professional lives, 42\% seek psychological treatment, and 51\% have contemplated suicide~\cite{ccri2014revenge, franks2015drafting}. Many face ongoing harassment, including the exposure of their personal information and intimate media online. Such actions not only perpetuate their trauma but also lead to a profound sense of powerlessness against the continual resharing of their content~\cite{citron2020new,ccri2014revenge,eaton20172017,citron2022fight,mcglynn_its_2021}. 

Addressing NCIM through legal channels has proven challenging, with systemic biases often discouraging victim-survivors from pursuing lawsuits. Many victims refrain from reporting due to past failures of systemic protection and potential for further harm~\cite{brodsky2021sexual, smith2014institutional, citron2022fight}. The financial burden of civil lawsuits further restricts access to justice, leaving many without recourse. The emotional and social impact of NCIM deeply affects how victims are perceived and treated across various societal dimensions, including in employment and legal settings.~\cite{sambasivan2019they, citron2022fight,ccri2014revenge,eaton20172017}.

\vspace{-0.5em}

\section{Related Work} \label{rr}

Next, we review NCIM scholarship and related CSCW research in greater detail. Despite overlap with established sociotechnical research domains like content moderation, online harassment, and digital consent, current research often falls short in tackling the unique challenges presented by NCIM. Likewise, NCIM work in law, privacy, psychology, usable cybersecurity, and communication offers important insights; however, they usually have limited engagement with NCIM's technical underpinnings. We outline these gaps as significant opportunities for addressing NCIM as a sociotechnical issue. 

\vspace{-0.8em}

\subsection{Legal lenses of NCIM}

\ud{A few prominent U.S.\ legal scholars have strongly advocated for more robust protections against the non-consensual sharing of intimate content. This advocacy has contributed to the proposal and enactment of NCIM laws at the state level in nearly all states~\cite{chen2016one,citron2020new, speier2022exploitation,bloom2014no}. Legal protections stipulate that the victim must be identifiable from the material, the material was intended to remain private, and that consent for its distribution was not given~\cite{citron2022fight}. Dissemination of sexually explicit material is typically a misdemeanor if the image was taken or accessed consensually but then distributed non-consensually. If the images were obtained non-consensually, criminal charges could also include a felony for unlawful surveillance. }More recently, Congress established a private right of action, enabling individuals whose intimate images were disclosed non-consensually to file a federal lawsuit against the discloser~\cite{coy2024violence,burnett2023criminalizing}.

Although these laws mark significant progress in combating NCIM, there are still substantial hurdles. Laws in a vacuum cannot immediately halt widespread dissemination, prevent personal information from appearing on abusive websites, or assuage emotional trauma. Civil lawsuits are very expensive, prohibitively so for those who do not have ready access to lawyers and high retainer fees. In criminal cases, victim-survivors may also not trust the legal systems in the domain of sexual harms~\cite{sambasivan2019toward,brodsky2021sexual,citron2020new}. Beyond these laws, the primary alternative for redress is the Digital Millennium Copyright Act (DMCA), which enables copyright owners to request the takedown of infringing content~\cite{citron2020new, goldman2020section230}. \ud{However, this approach is only viable if the victim took the photo themselves and can prove it---for instance, by providing a link to a publicly accessible version of the content, a method more suited for creative works than to intimate content. It also requires that the website be responsive to copyright takedown requests; many are not. }

\subsection{Empirical studies of NCIM}

Empirical psychology studies, including surveys and interviews, offer crucial insights into the prevalence and impact of NCIM~\cite{walker_systematic_2017}. Psychological research has explored the extent of victimhood---how common NCIM is~\cite{patel_prevalence_2022}, the impacts of this violation on mental health~\cite{walsh_if_2022, bates_revenge_2017}, and the factors that contribute to the perpetration of NCIM~\cite{schokkenbroek_receive_2023, clancy_dark_2019}.

Communication and gender studies scholarship frequently apply feminist theories to analyze NCIM as a form of gendered violence, aligning it with other abuses targeted at women and gender minorities~\cite{buiten2020s,naezer2021only, hearn2019my}. Previous research has detailed the range of harms experienced by victim-survivors, including emotional, economic, practical, social, and psychological impacts~\cite{henry_beyond_2015, dodge_nudes_2019, gius2022addressing, sambasivan2019they}. Studies have also examined attitudes towards NCIM, particularly people's allocation of blame between victim-survivors and perpetrators~\cite{hearn2019my}. \ud{Research has also addressed the factors leading to NCIM perpetration. }Content analyses of revenge porn websites have highlighted the significant role of the performance of masculinity in the illicit sharing of content~\cite{dekker2019don}, as well as reporting mechanisms for \ud{NCIM on technological platforms~\cite{de_angeli_reporting_2023}.}

Despite the valuable insights provided by these studies, they often are limited to the human/social/societal aspects of the issue. Moreover, while they offer detailed descriptions and analyses of the problem, there is a notable lack of focus on developing solutions for victim-survivors. This gap highlights the opportunity for research that interrogates technological strategies to mitigate NCIM's impact.

\subsection{Challenges in addressing NCIM}

\subsubsection{Privacy and cybersecurity}
Privacy and cybersecurity research characterizes NCIM as a significant privacy challenge. Previous studies have examined safe sexting practices~\cite{geeng2020usable,coduto2024delete}, attitudes toward security in intimate partner sharing~\cite{park2018share}, and defenses against intimate partner violence~\cite{tseng2020tools,chatterjee2018spyware,freed2019analyzing}, such as cyberstalking and blackmailing. This body of research underscores the complex social dynamics involved in sharing intimate content and managing threats like intimate partner violence~\cite{freed2018stalker, lee2019exploratory,tseng2020tools}. 

The concept of contextual integrity in information management suggests that once information is shared, individuals become co-managers of that information~\cite{nissenbaum2004privacy}. From this perspective, NCIM represents a mismanagement of co-managed information. Privacy research has also led to the development of systems designed to enhance visual privacy online. This research includes managing social media photos to obtain consent programmatically~\cite{olteanu_consensual_2018}, photography in public spaces~\cite{bo_privacytag_2014}, and preventing unwanted screen photography, screenshotting, and screen-recording~\cite{zhu_automating_2017, zhang_kaleido_2015, gu_anti-screenshot_2023,bai_fast_2023,jiang_information_nodate}. 

Despite these advancements, research specifically addressing the unique concerns of NCIM remains scarce. Traditional computer science and security approaches often overlook issues related to misogyny and the abuse of women and girls~\cite{tseng2022care,walker_systematic_2017}. To address this gap, our work focuses on the specialized privacy needs of victim-survivors when using technologies, exploring how better-designed systems can support their rights and well-being.

\subsubsection{Content moderation}

Online social platforms are a central locus point for NCIM harms, including doxxing, harassment, and privacy invasions~\cite{citron2014criminalizing,bates_revenge_2017,ccri2014revenge}. While Social Computing frequently speaks to online harassment and content moderation, current policies and practices often fail to adequately address NCIM issues.

Research has shown that platforms struggle to protect users from online harassment, with policies often being inconsistent and responses vague~\cite{jhaver_online_2018, blackwell_classification_2017}. Implementing generic solutions has proven impractical, leading many users to distrust the platforms' ability to effectively manage online harassment~\cite{schoenebeck_drawing_2021,schoenebeck_youth_2021}. In response, individuals and communities have developed various strategies to mitigate abuse. These include creating collective blocklists~\cite{jhaver_online_2018, geiger_bot-based_2016}, classifying and labeling harassment~\cite{blackwell_classification_2017}, sharing anti-harassment tools through networks~\cite{meisner_networked_2023}, employing auto-moderators or third-party bots~\cite{uttarapong_harassment_2021}, and enhancing the visibility of community norms~\cite{matias_preventing_2019}. Despite these efforts, managing online harassment remains a significant emotional burden, particularly for marginalized groups~\cite{uttarapong_harassment_2021}.

However, despite the critical importance of content moderation for NCIM, there is a notable absence of studies on how content moderation are applied to NCIM in reality.

\subsubsection{Technical interventions for consent}

In CSCW research, consent is crucial for developing systems that support sexual intimacy and well-being~\cite{strengers_what_2021, kannabiran2018design, nguyen_challenges_2020, zytko_online_2023,vaughn}. Consent in technology is also vital for ensuring agreement, access, and autonomy among users, their data, and other entities in the digital landscape~\cite{lee_building_2017}. However, existing systems addressing non-consensual content are limited in both scope and capabilities. Currently, the few systems in place primarily utilize hashing algorithms to focus on a subset of content types, including copyright infringement and child sexual abuse material. These algorithms convert content into a unique string, functioning as a one-way identifier. This means that while anyone can generate the hash from the original content, reconstructing the original from the hash is nearly impossible. This technique is widely used on platforms like YouTube and Twitter to identify and prevent copyright infringement. As another example, one well-known system, PhotoDNA, is used by many platforms to detect and block the distribution of illegal or unauthorized online content, including terrorism and child pornography~\cite{farid2018reining}. 

StopNCII is a rare system developed specifically for NCIM. It is a hashing-based system initially adopted by Meta to block unauthorized intimate photos from being uploaded to their platforms. Although StopNCII represents a significant step forward, it is similarly severely limited by its dependence on platform participation. Other platforms like WhatsApp and Telegram have introduced features that alert senders of screenshots and prevent them, which may deter unauthorized content captures~\cite{safedigitalintimacy}. 

Despite these developments, systems to combat NCIM are limited. There are significant opportunities for system-building work to address NCIM at its roots. Specifically, we propose expanding on existing work to develop more robust systems tailored to the unique challenges of NCIM, enhancing both detection and prevention capabilities.

\section{The \sts{}}\label{stack}

We introduce the \sts{}, an analytic device spanning social and technical realms, which allows us to pinpoint how technological components contribute to social problems. The sociotechnical stack is an extension of the conventional ``technology stack'' or ``tech stack,'' which is an abstraction used to describe the components that build an application---from hardware at the bottom, through the operating system, to the user interface at the top. See Figure \ref{fig:sstfig} and Section \ref{stacklayers} for further details. The sociotechnical stack integrates technical artifacts with social impacts, where each layer, while achieving a technical goal, also facilitates subsequent social impacts given specific contexts. This framework is particularly crucial in contexts with high societal importance like NCIM, where each layer in the stack plays a role in perpetuating harm, illustrating how technical design decisions can have profound social consequences. 

The hierarchical nature of the ``stack'' metaphor suggests that solutions could be applied at various levels. Addressing an issue like NCIM at a lower layer---such as hardware---can potentially resolve it at higher layers like the network or user interface. This holistic view encourages comprehensive solutions that move across the stack. As a conceptual framework, the sociotechnical stack is adaptable to specific system designs. For instance, in distributed computing, the stack might emphasize the network's role in task completion. In a decentralized architecture, the stack might focus on enhanced user control over content by allowing users rather than platforms to decide on content removal. The decentralized approach could empower users, or communities of users operating as nodes, to manage intimate content but might complicate efforts to control illicit redistribution. At any level and with any technical design choice, the sociotechnical stack lens demonstrates the need to tailor technical stack components to prioritize social impacts on users. 

The sociotechnical stack concept can extend beyond NCIM to address other social computing challenges where important social impacts are shaped and mediated by multiple layers of the stack. Take the problem of disinformation as a case study. Disinformation is information that is deliberately false or misleading, often perpetuated as a campaign for political purposes \cite{wilson2020cross, freelon2020disinformation}. A sociotechnical stack analysis could identify how layers of the sociotechnical stack enable the sharing and spread of disinformation. At the hardware level, there could be provenance of content capture and distribution (e.g. a photograph taken at a protest). At the storage level, there could be databases of prominent disinformation campaigns that would enable cross-platform collaboration. As with NCIM, a substantial challenge is in developing criteria and policies to agree on what qualifies as disinformation \cite{wilson2020cross}. At the network and algorithm levels, there could be trusted or trustworthiness protocols to deter well-known disinformation campaigns (e.g. use of Russian trolls in American race and politics) \cite{freelon2020disinformation}). At the application level, there could be cross-platform coordination in content moderation, rather than requiring that each platform detect and verify disinformation on its own, an expensive task. Finally, at the user interface level, there should be procedures to deter sharing and internalizing of disinformation content, while preserving speech and privacy rights. 

\begin{figure}[ht]
    \centering
    \includegraphics[width=\linewidth, trim={50 200 800 0}, clip]{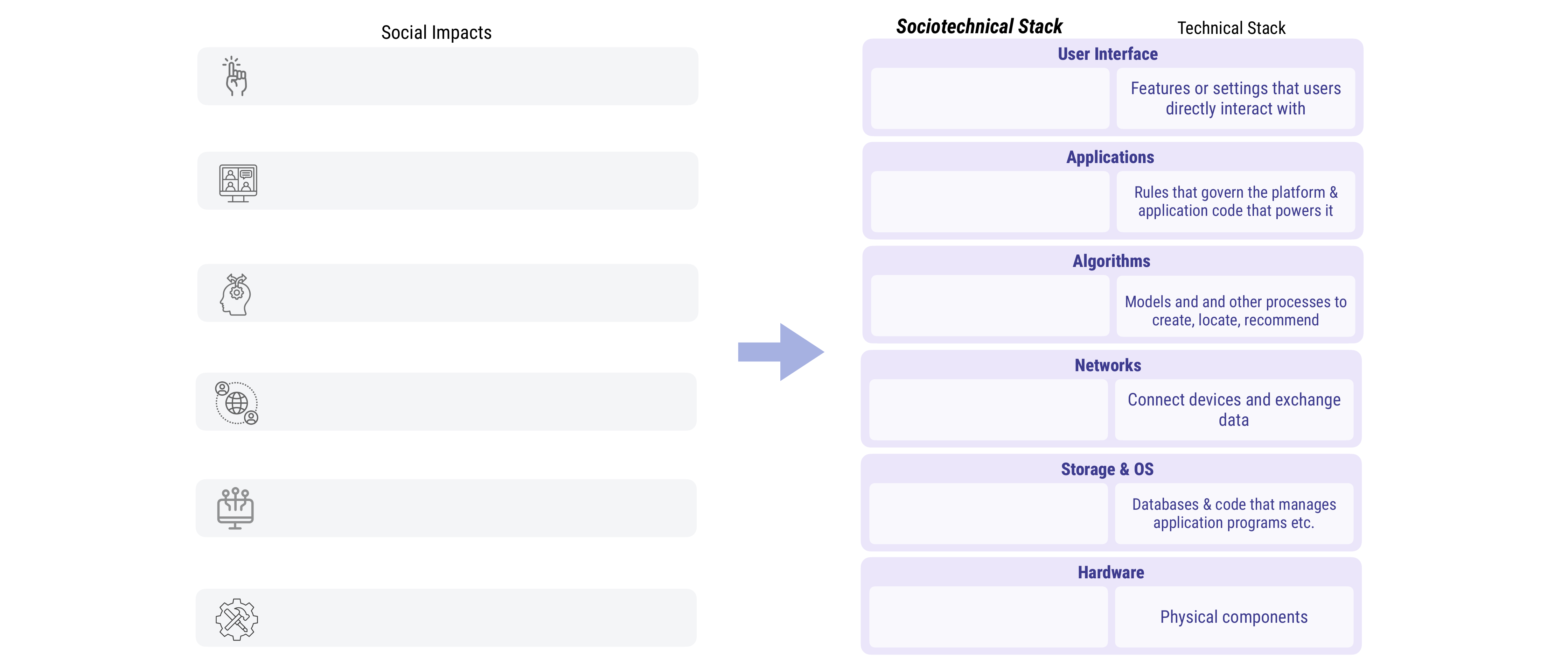}
    \caption{The conceptual framework \emph{sociotechnical stack} maps social impacts to the layers of the technical stack that facilitates or enables them. We share \href{https://osf.io/3q79x/?view_only=44b59c704d8a47bbb054438bd9c9d7ce}{downloadable versions} in PowerPoint and Keynote.}
    \label{fig:sst_generic}
\end{figure}

A sociotechnical stack analysis could be applied to other problem domains, ranging from bias in large language models to building accessible interfaces to surveillance with facial recognition technologies. The sociotechnical stack approach is useful for domains with large-scale social impacts that are shaped by design choices up and down the stack. Though this paper focuses on the problem of NCIM, Figure \ref{fig:sst_generic} provides a template for a visual sociotechnical stack analysis in other domains.\footnote{A link to the template for reuse is available \href{https://osf.io/3q79x/?view_only=44b59c704d8a47bbb054438bd9c9d7ce}{here} for PowerPoint and Keynote.}

\section{Data Sources}\label{source}

Narratives provide critical insights into the technical mechanisms that facilitate and perpetuate NCIM. To gather evidence and deepen our understanding of NCIM, we review prior works for firsthand accounts of victimization, utilizing interviews, quotes, and vignettes. We then employ a reflexive inductive thematic approach to explore \textit{sociotechnical} themes in these narratives, aiming to connect the technological components with the lived experiences of victim-survivors. We address the significant gap identified in previous sections: the lack of focus on the technologies that enable these violations~\cite{walker_systematic_2017}. We address this gap by analyzing how technologies translate into the lived harms of victim-survivors, posing questions such as ``How did technology facilitate the emergence of these experiences?'' and ``How did technology itself deliver these harms?'' 

To develop our vignette collection, we performed a literature review focusing on qualitative research related to non-consensual intimate imagery (NCIM). Using the search terms ``image-based sexual abuse'', ``online sexual abuse'', ``sexting'', ``revenge pornography'', ``intimate partner violence'', and ``non-consensual intimate imagery'', we searched ACM Digital Library, ProQuest Research Library, PsycINFO, and Communication \& Mass Media Complete databases. The first author selected studies rich in interview data and first-person accounts that detail the impact of NCIM, prioritizing narratives that provide deep and personal insights into harms. This includes interview studies that documented direct quotes from victim-survivors, and reports that contain stories and quotes from victim-survivors. We excluded the majority of legal reviews and quantitative studies due to the lack of focus on personal narratives. \textcolor{black}{We developed hierarchical codes of the interview text in the published articles. The hierarchical codes later became the themes detailed in Section 6. }The selected articles were reanalyzed to highlight \textit{how specific technologies facilitated and created these narratives and harms}. The result is a focused yet rich sample of victim-survivor experiences that illustrate how technology directly enabled NCIM. The sources for the vignettes and quotes are:
\begin{itemize}
    \item An interview study of 75 victim-survivors' experiences with NCIM~\cite{mcglynn_its_2021, henry_image-based_2019}
    \item A report drafted for legislators for stronger NCIM laws, including a survey on 361 victim-survivor experiences conducted by the Cyber Civil Rights organization~\cite{ccri2014revenge, franks2015drafting}
    \item An interview study of 18 victim-survivors focused on mental health~\cite{bates_revenge_2017}
    \item An interview study with 18 adults who have experienced sextortion~\cite{walsh_if_2022}
    \item A report on activists' and sex workers' experiences in social media content moderation~\cite{blunt2020posting}
    \item A report containing deepfake abuse survivor testimonies~\cite{compton2024deepfake}
    \item An interview study on sex-workers' privacy management strategies~\cite{mcdonald_its_nodate}
\end{itemize}

\subsection{Vignettes}

This section contains vignettes and quotes to illustrate the interplay of the sociotechnical stack's layers, enhancing our understanding of how each component contributes to NCIM harms. Vignettes are selected for their ability to explore these interactions in depth and serve as comprehensive illustrations of the sociotechnical stack in action. Vignettes draw out connections between \textit{multiple} layers, offering a holistic view of the sociotechnical dynamics at play. They serve as singular, emblematic victim-survivor experiences and will be referred to by their icons and will be used in subsequent analysis in Sections \ref{stacklayers} and \ref{roadmap}. Technology interactions are highlighted in the vignettes through \ul{underlining} to emphasize these elements. Quotes in subsequent sections will provide more direct mappings to \textit{specific} layers of the sociotechnical stack. 

\vspace{1em}
\noindent\textbf{\faEnvelope{}\,} \,\,\ud{``She and her ex-boyfriend had exchanged intimate photos throughout their three-year relationship, but she had never thought that he would use them to destroy her life \dots her pictures were \ul{on over 200 websites} and she had been \ul{inundated with unwelcome sexual propositions from men who had seen them}. The pictures had also been \ul{sent to her boss and a co-worker}. [She] spent the next few months trying to explain the situation to her employer, her family, her friends, and colleagues, and to \ul{plead with porn sites and search engines to remove her material}. Even after removing the photos, \ul{within two weeks her material was up on 300 websites}. At that point, [she] gave up trying to change her search results, and started the process to change her name. She couldn’t see any other way to escape the material that was \ul{following her everywhere}, jeopardizing her career, her psychological health, and her relationship''~\cite{ccri2014revenge}.}

\vspace{1em}

\noindent\textbf{\faVideoCamera{}\,} \,\,``Her ex-husband brought her to a hotel room and drugged her. She had no memory of what happened, but later found out that he and another man had raped her. She divorced him shortly after, and seven months later he \ul{sent a video containing footage of the rape} to the school board where she worked. She was fired from her job as a school superintendent immediately after the \ul{video was sent to her colleagues}, and the police were still investigating the rape at the time of the interview.''~\cite{bates_revenge_2017}

\vspace{1em}

\noindent\textbf{\faGlobe{}} \,\,``[She] found \ul{deepfake videos} of herself on various \ul{4chan threads}, alongside \ul{information about her college, name, and local area} \dots At first [she] started receiving a bunch of \ul{Snapchat notifications from people adding me} \dots and someone replied that \ul{someone posted my Snapchat on 4chan.}''~\cite{compton2024deepfake}

\section{NCIM Along the sociotechnical stack} \label{stacklayers}

This section examines how each layer of the sociotechnical stack enables NCIM harms and shapes the experiences of victim-survivors. To illustrate the complex interactions and downstream social consequences at each layer, we incorporate victim-survivor quotes and vignettes from sources mentioned in Section \ref{source}. 

Interactions within the sociotechnical stack require all layers to work in concert, much like how searching an online catalog involves the user interface, network instruction delivery, and database queries. Similarly, harms facilitated by the sociotechnical stack are distributed across its multiple layers. Thus, the focus of our analysis is not to isolate harms to a specific layer but to \textit{identify key technical components} within each layer that lead to specific social outcomes. We aim to highlight how changes at any single layer can influence the entire system, potentially mitigating or exacerbating NCIM harms. This approach allows us to provide a comprehensive view of the sociotechnical dynamics at play, offering insights into how technological design and implementation can be altered to reduce the incidence and impact of NCIM.

\subsection{User interface} 

The user interface (UI), as the top level of the technical stack, enables direct interaction and manipulation by users. It encompasses the visible elements of applications and includes features and settings that users can adjust. Crucially, the UI can facilitate NCIM harms by enabling loss of control over intimate media once sent, setting social media profiles to public by default, and allowing visible social networks to be used for malicious purposes.

\subsubsection{Loss of control over content after sent}

Online sharing of media results in the transfer of possession to the recipient, who can permanently save the content through various methods such as screenshots or screen recordings~\cite{shore_platform_2023,qin2024did,qiwei2024feminist}. Sharing intimate media, however, requires careful consideration of trust and reciprocity, as senders must navigate intimacy and negotiate trust~\cite{geeng2020usable}. This aspect of UI design often lacks mechanisms to prevent the unauthorized sharing of intimate media, despite the trust placed by the sender. A poignant example is provided by a victim-survivor's testimony:
\begin{quote} 
    \textit{``I sent this photo of my breast to this particular person. And they asked for more and more and more. I sent these pictures ... to this guy who I'd been seeing for months and months ... That's when he told me that he showed three of his friends what I'd sent him.''}~\cite{henry_image-based_2019} 
\end{quote} 
This quote underscores the ease with which trust can be violated and the profound impacts of such breaches, highlighting the need for enhanced protections.

\subsubsection{Visible social networks}

Victim-survivors often see their social networks used against them. On many platforms, social networks are publicly visible. Perpetrators exploit these visible connections, using publicly available information to tarnish reputations. The following account illustrates the misuse of visible social networks for intimidation and harassment:
\begin{quote} 
    \textit{``He told me he would take what [intimate photos] he had and show all my friends and put them on MySpace and all that and send them to other people in town.''}~\cite{walsh_if_2022} 
\end{quote}
This narrative not only highlights the direct threats faced by victim-survivors but also reflects a broader theme of exploitation that permeates many NCIM incidents, where personal and professional networks are abused to extend harm. Similarly, the perpetrator in \faEnvelope{}\, emailed her intimate photos to numerous contacts within her social and professional networks; \faVideoCamera{}\, had videos sent to her work colleagues.

\subsubsection{Default publicness}

Default publicness in social media design privileges ``a heteronormative assumptive posture of public expression''~\cite{cho2018default}. This makes personal information easily accessible and users readily reachable, sometimes directly enabling harassment. Online abuse among NCIM victim-survivors is prevalent and widespread; forty-nine percent report being harassed or stalked online by users who have seen their material~\cite{ccri2014revenge}. As a result, many victim-survivors are compelled to create new online identities to escape harassment and feel safe~\cite{henry_image-based_2019}. The public display of even basic personal information poses a safety threat~\cite{mcglynn_image-based_2017,mcglynn_its_2021,boyd2010social}. This exposure often leads to anxiety among victim-survivors, who fear unsolicited contact such as \textit{``Getting messages on Facebook from people I’d never met, telling me I had a nice body''}~\cite{henry_image-based_2019}. Many view subsequent online experiences as threatening, further isolating NCIM victim-survivors~\cite{mcglynn_its_2021}. The anxiety about being contacted via social media is profound, leading some to withdraw completely:
\begin{quote}
    \textit{``I don’t really have an online presence anymore because I’m just so terrified of \dots people contacting me because they’ve seen these pictures.''}~\cite{mcglynn_its_2021}. 
\end{quote}
These accounts highlight the need for a critical re-evaluation of privacy and consent mechanisms within UI design to better protect individuals from the far-reaching consequences of NCIM.

\subsection{Applications}

The application layer of the sociotechnical stack includes video hosting sites, social media platforms, mobile applications, and online forums. It consists of the rules, policies, and code that enable these platforms~\cite{lessig2009code}. This layer can facilitate NCIM harms through mechanisms such as impersonation and by providing a space for knowingly hosting non-consensual content.

\subsubsection{Impersonation}

The lack of stringent account verification on platforms facilitates abuse, including the creation of fake accounts, impersonation, and falsified posts. These tactics can be used maliciously to harm others, as illustrated by the following case:
\begin{quote}
    \textit{``They were a couple and she sent him photos, and then when they split up he set up a Facebook account with her name and he uploaded the photos to screw with her \dots from here and from the surrounding towns, it always tends to become public knowledge.''}~\cite{pavon-benitez_my_2022} 
\end{quote}
This quote highlights how social media platforms can be exploited to falsify identities, leading to significant social damage and public humiliation.

\subsubsection{Abusive platforms}
Platforms such as Reddit, Discord, 4chan, and Telegram are known to host forums and groups that are notorious for creating and sharing non-consensual sexual content~\cite{compton2024deepfake}. These platforms often serve as venues for the distribution of NCIM, exploiting their relative anonymity and community features. \faGlobe{} involve these platforms, showcasing how they facilitate the sharing of non-consensual content. Traditional content moderation techniques on larger platforms often prove insufficient, as harmful communities may simply migrate to less moderated platforms to continue their activities unchecked. This migration poses significant challenges in controlling the spread of NCIM and underscores the need for robust moderation practices across all platforms, not just the major ones~\cite{timmerman_studying_2023}.

\subsection{Algorithms}

Algorithms include generative AI models for creating new content, recommendation algorithms that inadvertently promote non-consensual content, and search engine algorithms that facilitate access to abusive material.

\subsubsection{Deepfakes}

2023 saw more deepfakes created than in all previous years combined~\cite{compton2024deepfake}. The advent of ``nudify'' models and their widespread availability have made it easier for perpetrators to generate and disseminate harmful content. The following quote illustrates the profound distress experienced by an individual whose altered images were circulated:
\begin{quote}
    \textit{``People were messaging me because they found me on a porn website \dots my name and my face were on there \dots my school and my hometown were also on there \dots oh my God, these videos had thousands of views on them. Hearing that the next time someone's in your hometown, they're going to fuck you. It's just terrifying.''}~\cite{compton2024deepfake}
\end{quote}
This experience highlights the severe social repercussions of deepfakes, demonstrating how quickly and extensively damaging content can spread, impacting personal security and mental health.

\subsubsection{Content recommendation algorithms}

Content recommendation algorithms aim to maximize views, likes, and shares. However, without mechanisms to discern consensual from non-consensual content, these algorithms can inadvertently distribute harmful material widely, a frequent issue on both mainstream and adult platforms. No safeguards currently prevent the circulation of content that may constitute sexual abuse. For NCIM victim-survivors, this distinction marks the difference between isolated harm and pervasive abuse. 

These algorithms also compromise privacy by recommending accounts. Consider the experience of a sex worker who tried to maintain separate personal and professional:
\begin{quote}
    \textit{``I wanted a second account with Facebook [for clients to interact with me]. I had a different email address \dots I didn’t want my friends to see it at all, but they were suggested to me [by Facebook] immediately \dots [so] I just deleted it right away.''}~\cite{mcdonald_its_nodate}
\end{quote}
This quote illustrates the privacy challenges individuals face due to the invasive nature of content recommendation algorithms, which can inadvertently expose sensitive aspects of a person’s life to unintended audiences, leading to potential social and professional harm.

\subsubsection{Search engine algorithms}

In \,\faEnvelope{}\,\, the victim-survivor had to undertake the laborious process of manually requesting search engines to deindex abusive content featuring her. This example illustrates the critical need for more proactive and victim-centered approaches in search engines to prevent the easy accessibility and perpetuation of abusive material online.

\subsection{Networking and the internet}

Networks connect devices, making online content potentially permanent due to widespread duplication.

\subsubsection{Near-infinite replication and spread}

Online content is duplicated for reasons both malicious and benign. Websites are often archived or automatically cached. Data scrapers harvest content from sites such as Reddit to boost traffic. Such practices underscore the challenge of online permanency: once posted, content may never be fully removable.

The internet is involved in most NCIM cases, as images are distributed one-to-one on messaging applications or broadcast to a wide audience~\cite{bates_revenge_2017}. Completely removing content from the internet is nearly impossible, a fact of which victim-survivors are painfully aware. These statements poignantly capture the enduring nature of online content: 
\ud{
\begin{quote}
    \textit{``There will never be a day in my entire lifetime that all the images of me could ever be deleted \dots It's a crime that doesn't just happen and then that's done. It's something that is continual, and this could continue for I don't know how long.''~\cite{mcglynn_its_2021}}
\end{quote}
}

While victim-survivors strive to control the spread of NCIM, they also grapple with the reality of the internet's permanence. A significant 54\% fear discovery of the material by their children or future descendants~\cite{ccri2014revenge}, leading to profound emotional distress and ongoing trauma.
\ud{
\begin{quote}
    \textit{``[It's become] a hidden obsession to always check my phone...and that’s kind of become the way that I’ve coped with it, constantly checking my phone, to the detriment of my work.''}~\cite{henry_image-based_2019} 
\end{quote}
}

\subsubsection{Broadcasting}
Websites facilitate the broadcasting of abusive materials. For instance, a perpetrator created new websites specifically to harm his ex-girlfriend.
\begin{quote}
    \textit{``Her ex-boyfriend posted an eBay auction for a disc containing naked photos of her, which she successfully had taken down from eBay. However, a year later, he created a porn website with the naked photos of her, which included her full name, the name of her town, the name of the college she taught at.''}~\cite{bates_revenge_2017} 
\end{quote}
Despite efforts by victim-survivors to remove content, media is frequently re-uploaded. Even with some measures in place on eBay to address abuse, the victim-survivor could prevent neither doxxing nor the public display of the content she attempted to remove. The inherent capabilities of the internet make it a challenging domain to control~\cite{lessig2009code}. Victim-survivors often feel helpless and trapped, tormented by the knowledge that their images are ``out of [their] control'' and ``hanging over'' them~\cite{mcglynn_its_2021}.

\subsubsection{Data provenance}

Data provenance concerns the origin, migration, and current location of data, posing critical questions for victim-survivors addressing privacy violations. Individuals often find themselves involuntarily playing a ``guessing game,'' forced to construct complex mental models to discern which social interactions might be safe. This necessity arises from their need to manage potential negative perceptions among those who may have seen their sexual images, as they ``hyper-analyze all of their social interactions''~\cite{mcglynn_its_2021}. The misuse of data extends into the realm of AI models, where non-consensual media are included in the training sets. This misuse initiates a new cycle of victimization, with these images being repurposed to develop deepfake models that facilitate further abuse.

\subsection{Storage and OS}

\subsubsection{Files}

A fundamental challenge with digital privacy begins at the level of file design and storage systems, which currently do not support the revocation of consent. Sexting lacks the privacy typically associated with physical intimacy. Despite agreements to keep shared photos private, digital systems facilitate their unauthorized circulation and duplication. Many people desire their ex-partners to delete intimate content post-breakup, yet there's no foolproof method to enforce such deletion~\cite{coduto2024delete}. 
\begin{quote}
    \textit{``I guess it was the thing to do, like you share nudes with your partner or boyfriend, and we broke up \dots We didn't end on very good terms \dots he [sent the photos around the school] and everyone kind of got my photo during class.''}~\cite{henry_image-based_2019} 
\end{quote}
This quote from a victim-survivor illustrates the lack of control over digital content once shared, even within trusted relationships. The inability to enforce digital privacy after a relationship sours shows a critical vulnerability in digital consent, exacerbating the risk of NCIM when intimate photos are misused. In broader social interactions, consent is often implicitly negotiated and nuanced, subtleties that current files fail to protect~\cite{ackerman2000intellectual, lee_building_2017}.

\subsubsection{Price of storage}

Perpetrators of NCIM exploit cheap storage for blackmail and abuse~\cite{bates_revenge_2017,mcglynn_its_2021}. The cost of storing digital content, such as videos and images, is minimal. This affordability means that once someone obtains a copy of content, they are likely to retain it indefinitely. There are few incentives for deletion. Unlike digital storage, physical storage incurs costs. Storing physical photographs in a photo album will ultimately run out of space.

\subsection{Hardware}

Cameras play a crucial role in NCIM. NCIM may arise from recordings initially made with consent (intended to remain private) or during acts of sexual assault, as depicted in \faVideoCamera{}\,. In other situations, recordings are made surreptitiously, without the person's awareness and without other crimes being committed. For instance, one ex-boyfriend set up covert cameras around the victim-survivor's home to film her~\cite{bates_revenge_2017}. Nonetheless, individuals do not consent to the recording because they are not aware of it. Other times, victim-survivors consented to sexual acts but did not consent to recording:
\begin{quote}
    \textit{``I didn't realize there was another friend who had managed to video me \dots peeked in the door, stuck a phone or something. He filmed me performing a sexual act on the person and then put it on Facebook''}~\cite{henry_image-based_2019}. 
\end{quote}

These document violations of consent and the illicit usage of recording devices on a vulnerable and unaware person. 

\section{Roadmap for NCIM Research} \label{roadmap}

As the final contribution of this paper, we present a research roadmap for NCIM. This roadmap builds directly on the gaps identified in our previous analysis, particularly the need for more computing research. It draws on insights derived from the sociotechnical stack and the narratives captured in victim-survivor quotes.

We use the sociotechnical stack as a framework to identify and explore potential research areas to reduce the harms associated with NCIM. By systematically examining each layer of the sociotechnical stack, we propose targeted interventions that could diminish the technical capabilities that currently facilitate NCIM. This includes innovating privacy safeguards, enhancing content moderation frameworks, and developing more robust consent mechanisms within digital platforms. This research agenda aims to bridge the gap between technical potential and social necessity, focusing on transforming how social computing and related fields can respond more effectively to the challenges of NCIM. No technology will eliminate the harms of NCIM; rather, the intent here is to significantly reduce harms using sociotechnical approaches.

\subsection{Research in user interfaces}

\subsubsection{Communicating consent.}
The process of intimate sharing online often lacks explicit, codified expressions of preferences regarding the handling of content after the fact. More critically, \ud{even when senders stipulate that materials remain private between partners}, these preferences are frequently disregarded. A novel approach could allow senders to specify their preferences at the interface layer and incorporate protective mechanisms down the stack to enforce preferences at the recipient's end. How could people communicate those preferences, and how could they stay with the media as it moves through various systems?

Possible methods for designing this elicitation include participatory design, which centers on the needs and desires of the communities involved in the design process, and speculative design, which promotes creative and radical thinking~\cite{robertson2012participatory,auger2013speculative}. \ud{Studies at the intersection of human-centered computing and sexual health have explored how interactions can be designed to support sexual well-being~\cite{brewer2006sexual}, the mediation of consent through technology~\cite{zytko2021computer,im2021yes}, and the effects of social media on sex work and sex-positivity.\footnote{\url{https://sexpositivesocialmedia.org}} }Future research can consider the flexibility and contextual constraints of preference elicitation features. For example, once a privacy preference has been expressed at the UI level and is now being implemented in the application, how can a \textit{possible revocation} of prior consent be designed? 

\subsubsection{Limiting the attack surface.}
Security research~\cite{geeng2020usable,nissenbaum2004privacy} indicates that people use various strategies to limit the exposure of personally identifiable information during intimate sharing. There is a need to integrate these insights into the design of platforms to better support privacy. In the development of messaging platforms, designers may consider elements important to intimate sharing, such as fostering trust and enabling the creation of private channels~\cite{ackerman2000intellectual,powell_blurred_2014}. Additionally, introducing new social sharing modalities can provide users with more flexible, controlled, and secure options for both sending and receiving content.

\subsection{Research in applications}

\subsubsection{Content moderation and takedowns.} Content moderation encompasses a wide range of activities aimed at mitigating harmful content on platforms, including harassment and misinformation~\cite{gillespie2018custodians,kurt}. Key questions at the intersection of NCIM and content moderation involve identifying non-consensually shared content, supporting victims in requesting content takedowns, and establishing regulatory requirements for companies to manage NCIM effectively. While many websites have community guidelines that regulate user behavior, not all have specific policies addressing non-consensual content sharing, and even those that do may lack enforcement procedures~\cite{de_angeli_reporting_2023}. The process for victim-survivors requesting takedowns of NCIM content lacks transparency, leaving many unsure of the outcomes. 

Further research may study how content circulates through online communities. Misogyny and toxic masculinity are prevalent online and may intensify when combined with graphic and traumatic imagery~\cite{jones2020sluts}. Techniques such as nudging have shown effectiveness in fostering healthier behaviors in online settings, suggesting potential applications for deterring the spread of NCIM. However, research also indicates that awareness of the non-consensual nature of sexual content may \textit{heighten} toxic masculine behaviors and motivate further consumption~\cite{dekeseredy1993male,dekeseredy2016thinking}. This area demands careful consideration of the balance between social and antisocial behavior.

\subsubsection{Trauma-informed social computing.}
Those impacted by NCIM may be seeking recovery in a complex emotional state, requiring support that is trauma-informed~\cite{chen2022trauma}. Online experiences that were previously mundane may now become traumatizing. Research may explore how to design interfaces that minimize triggering experiences following trauma. This problem is not limited to NCIM victim-survivors, but extends to individuals who experienced a variety of traumatic events online, including but not limited to stalking, doxxing, intimate partner violence, and blackmailing~\cite{southworth2007intimate,freed_digital_2017}. \ud{Tseng et al.\ note that continuity is an important principle for designing interventions to address IPV; information, relationships, and approaches should be consistent when caring for IPV victims~\cite{tseng2022care}. This, combined with other types of internet-related trauma, suggests that design ought to account for usage after traumatic events~\cite{chen2022trauma,scott2023trauma}. }

\subsection{Research in algorithms}

\subsubsection{Hashing.} 
Hashing algorithms are essential in preventing the spread of NCIM. However, this is not without limitations. Platform participation is entirely optional, it may be challenging to detect altered images, and requiring victim-survivors to verify their identity as part of the process can further infringe on their privacy. Research should strive for more comprehensive solutions that are effective regardless of platform policies, while accounting for the very real possibility of such systems being misused themselves, e.g., by perpetrators of intimate partner violence wishing to eliminate evidence of abuse~\cite{freed2018stalker}.

\subsubsection{Deepfakes.}
The emergence of sexually explicit AI-generated content presents another dimension of NCIM harm. Studies indicate that deepfake sexual abuse constitutes a significant portion, accounting for 96 to 98\% of all deepfake content~\cite{kwok_deepfake_2021,karasavva_real_2021,compton2024deepfake}. This alarming prevalence, coupled with existing inadequacies in legal protections, underscores the urgent need for policy interventions. Deepfakes pose unique challenges that redefine the boundaries of consent in the creation of media. Unlike traditional forms of NCIM, deepfakes eliminate the need for physical proximity or explicit recordings, as perpetrators need only easily accessible benign images or videos from the victim's social media profiles. Exploiting default publicness and leveraging advancements in generative AI to produce nude images, deepfakes facilitate the creation and dissemination of NCIM without consent or awareness. The distinction between consent and violation in deepfake technology is stark. Unlike cases where intimate media was initially shared consensually with a partner, deepfakes leave little room for ambiguity. While not inherently limited to non-consensual content, 
the most prevalent uses of deepfake technology are harmful, and necessitate a critical reassessment of its legal status and the protections afforded to individuals.

\subsection{Research in networks}

\subsubsection{CDNs.}
Content Delivery Networks (CDNs), such as CloudFlare, have the potential to recognize and refuse hosting NCIM. For example, CDNs may ensure that non-consensual material is not propagated across their servers. Given the central role of hosts in the online ecosystem---where a single host can support numerous websites---their influence is pivotal. This power necessitates caution; it grants hosts substantial control over decisions concerning platform content, potentially affecting how content is moderated or censored. It is important to prevent censorship of legitimate expressions while curbing the spread of harmful material.

\subsubsection{Consent protocols.}
Could protocols be utilized to enforce consensual data transmission? Could we then affirm that the transmission of data---particularly intimate content---is consensual? This approach could significantly reduce NCIM harms which primarily stem from unauthorized sharing over networks. However, implementing such a system requires careful governance to balance the deterrence of non-consensual sharing with the need to maintain the free flow of knowledge and other beneficial data. Developing media protocols including licensing agreements could help deter unauthorized use. One challenge here is applying these solutions to emerging decentralized platforms, where control and enforcement mechanisms are not as straightforward as in centralized systems. Research in this area may consider innovative approaches to protocol design that can operate effectively within different topologies.

\subsection{Research in storage and OS}  

\subsubsection{Deterring duplication.}

Unauthorized duplication of content is a central issue in many NCIM cases. There are several forms of unauthorized duplication of content, including screenshots replicating intimate details. This prompts the question: could files themselves be designed to embed mechanisms that verify and protect user consent, akin to how institutions manage document access on cloud services? Future research in protocols may address duplication prevention measures. Although it is impossible to \textit{fully} prevent duplication, introducing barriers and restrictions may mitigate some harmful practices: even if protections don't prevent \textit{all} harms, barriers may sufficiently deter from a \textit{portion} of non-consensual screenshots or recordings. Deterrence may even communicate senders' privacy preferences.

\subsection{Research in hardware} 

\subsubsection{Consensual cameras.} 
Is it feasible to ensure consent is respected by recording devices? For example, the MacBook webcam is designed with a green light that is hardwired to turn on whenever the camera is on.\footnote{\url{https://support.apple.com/en-us/102177}} This design helps prevent unauthorized recordings by malware and mitigates concerns about secret recordings. Meta Ray-Bans are designed with a continuously blinking capture LED light to signal when the camera is active. If the LED is obscured, the glasses will stop recording, prompting the user to clear the obstruction.\footnote{\url{https://about.meta.com/ray-ban-stories-privacy/\#built-for-your-privacy}} The LED serves as a transparent indicator to ensure that people are aware when recording is taking place. Similarly, camera-equipped phones sold in Japan have a shutter sound that, by industry convention, is not possible to disable.\footnote{\url{https://www.engadget.com/2016-09-30-japans-noisy-iphone-problem.html}} This privacy protection was established to avoid surreptitious photos and filming. 

The direct wiring of the webcam light ensures the feature cannot be bypassed. Could the camera's hardware be programmed to interact with the application layer to ensure consent is obtained? Could future recording devices be designed to actively seek and respect the consent of the subjects before recording begins? Hardware solutions like this are particularly impactful because they operate at the lowest level of the technical stack, embedding privacy protections directly into the device itself. By implementing these features at a fundamental level, just as data flows up the technical stack, the benefits may flow up the sociotechnical stack.

\section{Conclusion}

NCIM is both a technical and sociotechnical challenge, interacting deeply with many social and computing elements. The roots of sexual abuse and the misogyny that drive NCIM are historical and pre-date the internet. The sociotechnical approach presented in this paper cannot \textit{fully solve} the underlying issues of abuse, nor fully address the shaming and gendered imbalances experienced by victim-survivors, it offers strategies to reduce the widespread proliferation of harmful content. This paper addresses the critical gap in how technologies not only facilitate but also generate NCIM at scale. Through the sociotechnical stack framework, we analyze the interplay between social impacts and technical elements, highlighting the role of features such as generative AI models in creating non-consensual content. Our approach demonstrated the practical application of the sociotechnical stack to dissect the complexities of NCIM, revealing how these technologies influence the creation, obtainment, and dissemination of content. We present our findings and outline a research roadmap of sociotechnical research in NCIM to support victim-survivors, deter perpetrators, and foster consent in digital interactions. 

\begin{acks}
    We thank our colleague Daniel Szalkiewicz for early feedback on this work. This material is based upon work supported by the National Science Foundation under Grants 1763297 and 2311102.
\end{acks}

\color{black}
\bibliographystyle{ACM-Reference-Format}
\bibliography{example}

\end{document}